\documentclass[draft]{agujournal2019}
\usepackage{url} 
\usepackage{lineno}
\usepackage{float}
\usepackage[inline]{trackchanges} 
\usepackage{soul}
\draftfalse

\journalname{}

\begin{document}

%
%


\title{A Global Analysis of Pre-Earthquake Ionospheric Anomalies}

%
%




\authors{Luke Cullen\affil{1, 2, *}, Andy W Smith\affil{1, 3, 4, *}, Asadullah H Galib\affil{1, 5}, Debvrat Varshney\affil{1, 6}, Edward J. E. Brown\affil{1, 2, 7}, Peter J Chi\affil{1, 8}, Xiangning Chu\affil{1, 8}, Filip Svoboda\affil{1, 2}}


\affiliation{1}{Frontier Development Lab, Sunnyvale, CA, USA}
\affiliation{2}{University of Cambridge, UK}
\affiliation{3}{Northumbria University, UK}
\affiliation{4}{University College London, UK}
\affiliation{5}{Michigan State University, USA}
\affiliation{6}{University of Maryland Baltimore County, USA}
\affiliation{7}{British Antarctic Survey, UK}
\affiliation{8}{University of California Los Angeles, USA}
\affiliation{*}{These authors contributed equally to this work}




\correspondingauthor{Luke Cullen}{lshc3@cam.ac.uk}




\begin{keypoints}
\item Statistical investigation of ionospheric density anomalies in the 12 hours before earthquakes using 20 years of data.
\item Statistically significant regional anomalies identified but no globally consistent signature.
\item Regional analysis ionospheric precursors should be considered as a key part of earthquake forecasting.
\end{keypoints}

%
%

%
%


\begin{abstract}

Local ionospheric density anomalies have been reported in the days prior to major earthquakes. This global study statistically investigates whether consistent ionospheric anomalies occur in the 24 hours prior to earthquakes across different regions, magnitudes, temporal and spatial scales. We match earthquake data to Total Electron Content (TEC) data from 2000-2020 at a higher resolution and cadence than previous assessed. Globally, no significant, consistent anomaly is found. Regionally, statistically significant ionospheric anomalies arise in the 12 hours prior to earthquakes with $p \leq 0.01$ following Wilcoxon tests. For the Japanese region we find a median negative ionospheric anomaly of around $0.5\,\mathrm{TECU}$ between 3 and 8 hours before earthquakes. For the South American region, the median TEC is enhanced by up to $\sim 2\,\mathrm{TECU}$, between 7 and 10 hours before an event. We show that the results are robust to different definitions of the ``local'' region and earthquake magnitude. This demonstrates the promise of monitoring the ionosphere as part of a multimodal earthquake forecasting system.

\end{abstract}

\section*{Plain Language Summary}

The density of the upper, ionized layer of Earth's atmosphere has impacts on communications and navigational equipment. We monitor this density through detecting interference in our navigational guidance systems. We use density data of this layer of the atmosphere to investigate whether there are any detectable and consistent signals emitted before earthquakes, that could be used with other indicators to help forecast earthquakes. When breaking down our analysis into regions, we find consistent and significant signatures in the 12 hours prior to earthquakes. In agreement with previous studies we find no globally consistent signature, which implies that anomalies are dependent on subsurface properties of the region analyzed. Our findings support a more regional analysis of pre-earthquake processes and that atmospheric disturbances should be considered as a valuable additional data source in future earthquake forecasting studies.

%
%

\newpage

\section{Introduction}

The ionosphere is one of the upper layers of Earth's atmosphere, located at altitudes greater than $\sim50\,\mathrm{km}$. It is an ionized region, primarily due to illumination with solar UV radiation, although precipitation of energetic particles from near-Earth space can also contribute on a local scale. Due to these ionizing sources, the density and composition of the ionosphere vary with time of day, latitude, and the levels of solar and geomagnetic activity \cite{Mendillo2006, Prolss2008, Kane2003}. This variability can impact the propagation of radio signals, interfere with communication systems, and degrade the accuracy of navigational tools. Due to this interference, we can remotely monitor ionospheric properties through their impact on GNSS (Global Navigation Satellite Systems) signals. Dual-frequency GNSS receiver measurements can be used to calculate the integrated electron density, known as Total Electron Content (TEC), along the line of sight to the spacecraft. This can then be converted into the vTEC, or vertical total electron content, to obtain a measurement of TEC directly above the location of the GNSS receiver.

While many factors impacting the ionosphere are external in origin (e.g. solar illumination \cite{Liu2009b}), it can also be impacted by subsurface events including earthquakes. For example, the magnitude 9.0 Tohoku earthquake in 2011 \cite{Rolland2011}, as well as the recent 2022 Tonga volcanic eruption \cite{Zhang2022}, had large and unambiguous impacts on the ionosphere in the hours after the events. \cite{Galvan2011} showed that two earthquakes of magnitude 8.1 and 8.8 (offshore Samoa and Chile respectively) were associated with fluctuations in TEC, inferred to be a result of the propagation of atmospheric gravity waves perturbing the ionosphere \cite{Hickey2009}. However, we note that the seismogenic signals detected by \cite{Galvan2011} after these two large tsunami-triggering events only had a typical amplitude of $\sim 0.1 - 0.2$ $\mathrm{TEC}$ Units $\mathrm{(TECU)}$, and thus required careful data processing to extract. \cite{Vijayan2022} assessed the impact of the non-uniform sampling and aliasing of ionospheric measurements on the detection of seismogenic ionospheric perturbations. They found that these inherent properties of the data must be considered to distinguish these perturbations from the background. Nonetheless, \cite{Astafyeva2019} demonstrated that the ionospheric impact of earthquakes can be seen at least as low as magnitude 7.4.

It has also been suggested that there may be changes in the ionosphere in the hours, days or weeks prior to an earthquake occurring \cite<e.g.>[]{pulinets1998seismic, Liu2009, Ikuta2022}, though we note that the physical mechanisms for such observations are as yet unclear \cite<see comprehensive review by>[]{conti2021critical}. Inspecting 20 earthquakes above magnitude 6 near Taiwan, \citeA{Liu2004} showed a statistically significant decrease in ionospheric TEC up to five days before an earthquake in the 1800 - 2200 (evening) local time sector. Later work has since shown that various TEC anomalies may be found prior to earthquakes, and that ionospheric TEC may be enhanced or decreased \cite{Liu2018, Ulukavak2020}, perhaps depending on the solar local time. We note that some works have suggested that issues arise when accounting for variable data quality, and the definition of a TEC ``anomaly'' often varies between studies. Nonetheless, it is widely agreed that searching for pre-earthquake signatures in TEC data shows promise \cite{Lim2019, Ikuta2022}.

\citeA{Thomas2017} completed a global study, concluding that there was no consistent, global TEC anomaly in the days before an earthquake.  However, they noted that anomalies could be localized or last only a few hours, which would have prevented them from being detected in their study. \citeA{Zhu2020} later looked at ionospheric disturbances up to 15 days before earthquakes, limiting their sample to those earthquakes that occurred inland (i.e. not sub-oceanic) due to GPS data coverage. Their study also concluded that there were little to no consistent signatures of TEC anomalies.

In this work, we use higher spatial and temporal resolution than has previously been considered to statistically assess whether any consistent changes in ionospheric TEC can be identified before earthquakes. We use a large historical dataset spanning 20 years, to explore both global and regional observations. We have chosen to make as few down selections of data as possible, and applied a simple, easily reproducible scheme in which we compare TEC values around each earthquake with observations made at a previous non-earthquake time. Such a scheme would be lightweight and straightforward to apply as a regional monitor. We show that such a lightweight monitor, comparing statistical parameters within regions of the ionosphere, may be suitable for identifying ionospheric anomalies that can precede earthquakes.

\section{Data}

This study uses data from the Madrigal database [cedar.openmadrigal.org], which provides global maps of vertical TEC, calculated from GNSS data \cite{Rideout2006}. These data are provided on a 1° by 1° grid at a temporal cadence of 5 minutes.  We note here that though global, the data are incomplete, with typical global completeness of $\sim 25\%$ \cite<e.g.>[]{sun2022matrix}.

To classify earthquakes and label times of events we use the USGS earthquake catalog \cite{usgs2022}. This catalog records critical features such as the event time, magnitude, epicenter location and epicenter depth of earthquakes.

To maximize data completeness while permitting the largest possible set of historical data, we use data between 2000 and 2020 as our statistical sample.

\section{Method}

Ionospheric TEC is influenced by a large number of contributing factors. These factors include solar illumination and incident energetic particle flux, meaning that TEC will vary annually, diurnally and stochastically depending on the current space weather conditions. Previous statistical studies have highlighted different time-scales of ionospheric perturbations due to seismic activity ranging from hours to days before earthquakes \cite{pulinets2003main, pulinets2004ionospheric, liu2006statistical, le2011statistical, li2013statistical, parrot2018statistical}. To probe the impact of seismic effects globally and across all available events we must standardise the time-scale observed. Overall, following case studies illustrated in figure \ref{fig: Example}, the largest variation between TEC values appears when comparing the 12 hours either side of an event, with the 24 hour period immediately before. This standardisation will allow for the detection of TEC variations taking place in the hours prior to earthquakes, as seen in some previous studies \cite{kelley2017apparent, muafiry20203}, but may obscure variations that are observed over days to weeks before events as seen in other studies \cite{parrot2018statistical}.

Furthermore, we do not remove events during geomagnetically active intervals \cite<c.f.>[]{Thomas2017}, relying instead on the large statistical dataset we are deploying, and the normalization method. This ensures that if any operational ionospheric monitoring used these procedures it would be agnostic as to the current ``space weather'', though the performance would likely degrade depending on conditions. Such considerations will form a future extension.

Due to data availability, as discussed above, it is not always possible to compare the TEC value at a specific location (e.g. $1^{\circ}\mathrm{x}1^{\circ}$ locale), but this is not necessary for our study. Instead, we evaluate the statistical properties (i.e. median) of a small region of the ionosphere around a central location, e.g. $\pm 3^{\circ}$ of longitude and latitude. This allows, first for a more regional analysis, and secondly to create a more representative TEC value base on the median of multiple measurements. We will vary the radius of the area used throughout the study to understand the spatial extent of the anomaly. It is known that absolute TEC values can display an error margin of 1-2 TECU due to satellite and receiver biases and differences in calculation methods used over time. However, this study is focused on the relative difference of TEC values that are close in time (within tens of hours) and space (within a few degrees), and therefore these errors should be minimized.  Further, we presume that these uncertainties would manifest as random noise in our sample, for which we will examine the statistical significance of any signal.

Each median observed TEC is compared to the median TEC exactly 24 hours previously using equation \ref{eq: Anomaly}. This is applied to all measurements in the 12 hours before and after each event. 

\begin{equation}
    \mathrm{TEC_{Anomaly}} = <\mathrm{TEC}>_{T} - <\mathrm{TEC}>_{T-24h}
    \label{eq: Anomaly}
\end{equation}

In equation \ref{eq: Anomaly}, $<\mathrm{TEC}>$ is the median of the TEC values measured within a defined region at time $T$. If less than two data points are recorded in the region with the time window in question, this epoch is ignored and the anomaly is not calculated due to a lack of data.

To determine the statistical significance of any change in ionospheric TEC we employ the Wilcoxon test \cite{Wilcoxon1945, Conover1971}. The non-parametric Wilcoxon test compares the distributions of the $\mathrm{TEC_{Anomaly}}$ at each epoch on the day of the earthquake and the corresponding epoch 24 hours prior, to assess whether the point-wise difference between the distributions is symmetric around zero. If it is asymmetric around zero we can infer that any deviation in the distribution of TEC anomalies observed before an earthquake is statistically significant. A priori we arbitrarily set an upper limit on significance at a p-value of $p = 0.2$, though we will report the calculated p-value associated with our results.

\section{Results}

The results from applying our TEC anomaly detection method will first present a case study of two major earthquakes, followed by the global and regional statistical analysis covering all events in the dataset. We will then present a more detailed analysis of the TEC anomaly properties for the Japanese and South American subduction zone regions, including magnitude and spatial extent.  We note that though we present the results for two representative regions, we have tested our method on 5 regions of major earthquakes, the results of which are included in the SI.

\subsection{Case Study Examples}

\begin{figure}
\includegraphics[width=\textwidth]{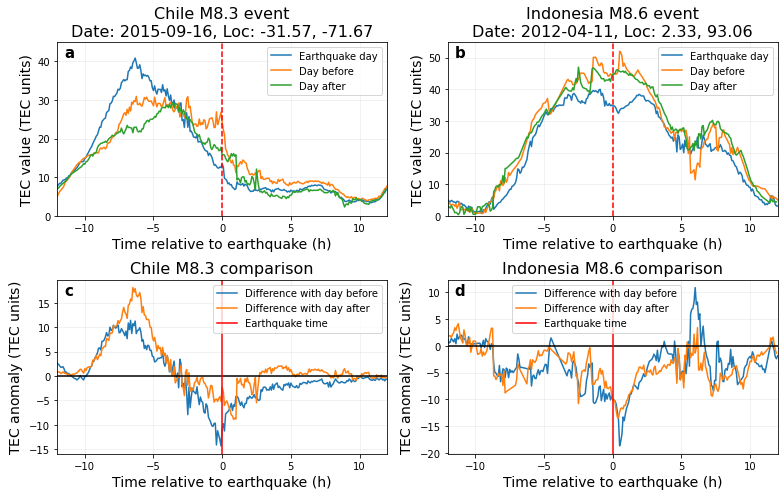}
\caption{TEC variation within $3^{\circ}$ of earthquake epicenters for a major Chilean and Indonesian event. a) and b) directly show the median TEC value for each event on the day of, the day before (- 24 hours) and the day after (+ 24 hours), each event. c) and d) show the resulting TEC anomaly for each earthquake day calculated by taking the difference between the median TEC on the day before, in blue, and with the day after, in orange.}
\label{fig: Example}
\end{figure}

Figure \ref{fig: Example} details two case study events, showing the variation in median TEC for the 24 hours centered on a 2015 Chile magnitude 8.3 earthquake (figures \ref{fig: Example}a, \ref{fig: Example}c) and a 2012 Indonesian magnitude 8.6 earthquake (figures \ref{fig: Example}b, \ref{fig: Example}d). The top row (figures \ref{fig: Example}a, \ref{fig: Example}b), shows the variation in TEC for the day of the earthquake and the two days on either side (i.e. $t \pm 24 \mathrm{hours}$), while the bottom row shows the resulting TEC anomaly, calculated as per equation \ref{eq: Anomaly}, for the day of the earthquake compared to the previous and following days.

The top row of figure \ref{fig: Example} shows quite large ($\sim 10\,\mathrm{TECU}$) differences between the TEC values recorded for the three days around both example earthquakes. For the 2015 Chilean earthquake, we can see that on the day of the earthquake the TEC appears to be enhanced between 4 and 10 hours prior to the earthquake and suppressed between 4 hours before and the time of the earthquake. This could perhaps be evidence of the solar local time dependence noted by other studies \cite<e.g.>[]{Liu2018, Ulukavak2020}, where TEC is suppressed or enhanced depending on solar local time of the observation. Meanwhile for the 2012 Indonesian earthquake TEC appears to be suppressed for several hours before and after the earthquake. These patterns are consistent irrespective of the comparison being made to either the previous or following days. However, it is unclear from these individual examples if the anomalies are caused by earthquakes, or stochastic processes such as space weather. A statistical analysis of many events will avoid this uncertainty by negating the events of individual space weather events.

\subsection{Statistical Analysis}

\begin{figure}
\includegraphics[width=\textwidth]{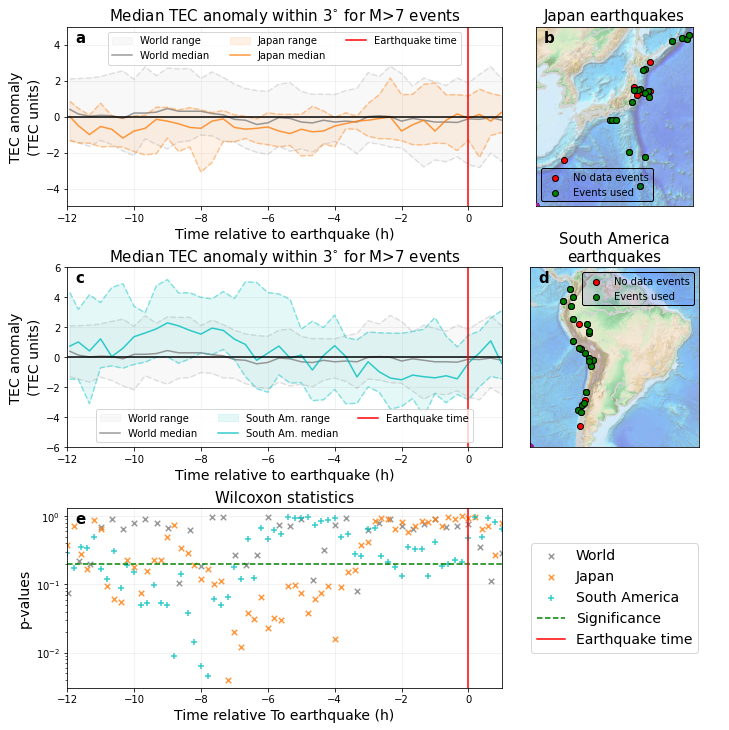}
\caption{Statistical analysis of TEC anomalies within $3^{\circ}$ of M$>$7 earthquake epicenters. The anomaly is a result of comparison with the preceding day, for 12 hours prior to earthquakes. a) Median TEC anomalies for events with sufficient data in the Japanese region shown in b). c) Median TEC anomalies for events with sufficient data in the South American region shown in d). e) Statistical significance of TEC anomalies for both regions covered in a)-d). Results for the global analysis are shown as a reference in the background of plots a) and c). See SI for North American, Tongan and Indonesian regional statistics.}
\label{fig: RegVar}
\end{figure}

Figure \ref{fig: RegVar} shows superposed epoch analyses of the TEC anomaly measured within $\pm 3^{\circ}$ of the earthquake epicenter, for earthquakes greater than magnitude 7, in the 12 hours before the events. Figures \ref{fig: RegVar}a and \ref{fig: RegVar}c show the median TEC anomaly, in solid lines, for the whole global dataset (in gray) and for two specific regions: Japan and South America in orange and cyan respectively. The interquartile range of the distributions are indicated with the shaded regions.  Figures \ref{fig: RegVar}b and d show the location of the earthquakes included in these statistical summaries noted in green, while those in red appear within our dataset but have insufficient ionospheric data to be included. For the Japanese earthquakes, 27 occur within our time interval, of which 20 have the requisite ionospheric data. Similarly, for the South American earthquakes: 25 occurred within our time interval, 21 of which have sufficient data for our analysis. Figure \ref{fig: RegVar}e shows the p-values obtained by the Wilcoxon test, with the a priori significance level of 0.2 indicated with a green dashed line.

From the gray distributions (replicated in figures \ref{fig: RegVar}a and \ref{fig: RegVar}c) we can see that for the full global selection of earthquakes within the 2000-2020 time interval, the median TEC anomaly hovers around zero. In contrast, when restricting the sample to those within the Japanese region (Figure \ref{fig: RegVar}a) the median TEC anomaly is below zero until 3 hours before earthquake time, with the upper quartile found to hover around zero, for 3 to 8 hours before events. Inspecting the results of the Wilcoxon test on these distributions of TEC anomalies (Figure \ref{fig: RegVar}e, in orange) we see that between 4 and 8 hours before the earthquakes the TEC distributions obtained are significantly below zero, with significance peaking at $p \leq 0.005$. Inspecting the results for the South American earthquakes (Figure \ref{fig: RegVar}c), the median TEC anomaly is enhanced between 6 and 10 hours before the earthquakes. The Wilcoxon test (Figure \ref{fig: RegVar}e, cyan) confirms the significance of this enhancement, with $p \leq 0.005$ approximately 8 hours before the earthquakes.

\begin{figure}
\includegraphics[width=\textwidth]{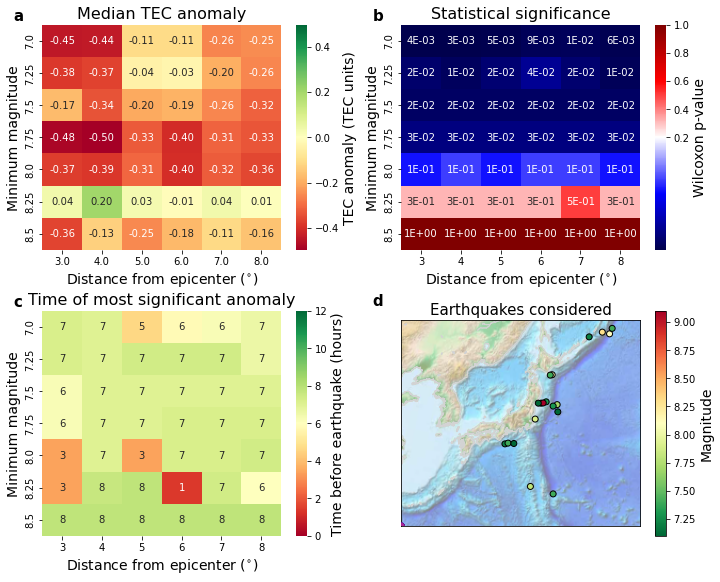}
\caption{Anomaly dependence on earthquake magnitude and distance from epicenter for Japanese earthquakes. Minimum magnitude implies that only events over the stated magnitude are considered. Support for magnitudes is as follows: M$>$7: 20,  M$>$7.25: 14,  M$>$7.5: 7, M$>$7.75: 6, M$>$8: 4, M$>$8.25: 2, M$>$8.5: 1. Distance from epicenter implies all data available within this radius are used to calculate median TEC values. a) Median TEC anomaly value at the time with the highest statistical significance within 8 hours prior to the earthquake time. b) Minimum Wilcoxon p-value for TEC anomaly within 8 hours prior to earthquake. c) Time before the earthquake with the most statistically significant TEC anomaly. d) Map of earthquakes used in the analysis.}
\label{fig: Var}
\end{figure}

Figure \ref{fig: RegVar} shows the statistics for a specific region around earthquakes ($\pm 3^{\circ}$) and for earthquakes larger than a given magnitude (greater than magnitude $7$). We now examine how our results depend upon these choices. Figure \ref{fig: Var} shows four panels examining these relationships using the earthquakes within the Japanese region. Figure \ref{fig: Var}b shows the p-value of the most significant epoch in the twelve hours before the earthquakes, according to the Wilcoxon test, as a function of minimum earthquake magnitude included in the sample and regional extent for inclusion of TEC measurements.  Here, as above, we have chosen a significance level of 0.2: values larger than this are colored red, while more significant values are blue.  Figure \ref{fig: Var}a shows the median TEC anomaly for the most significant epoch, with green colors indicating an anomaly greater than zero (i.e. an enhancement in TEC) and red indicating an anomaly below zero (i.e. a suppression of TEC). We note that the most significant epoch may not correspond to the largest median TEC anomaly as the significance of the distribution is tested, and not the magnitude of the median value. Figure \ref{fig: Var}c shows the time of the most significant epoch relative to the earthquake. Figure \ref{fig: Var}d shows the earthquakes within the region used for the analysis, colored by magnitude.

The domination of red in figure \ref{fig: Var}a) shows that earthquakes greater than magnitude 7 in the Japanese region see a reduction in the observed median TEC.  The median TEC anomaly is typically $\sim0.4\, \mathrm{TECU}$ within $3^{\circ}$ and reduces as the spatial region considered increases. As the minimum magnitude included in the sample increases (e.g. moving down Figure \ref{fig: Var}a) we see that the median TEC anomaly is relatively steady, though it decreases at a minimum magnitude of around 8.25. This is likely due to the number of earthquakes in the sample decreasing to just two events, which forces the Wilcoxon p-value above the significance threshold (figure \ref{fig: Var}b). It is possible from figure \ref{fig: Var} that as the minimum magnitude increases, larger median TEC anomalies are seen at greater distances from the epicenter. This observation holds moving from top-left in figure \ref{fig: Var}a to the square at minimum magnitude 8.0 and distance from the epicenter of 8$^{\circ}$. The distributions of TEC anomalies at the most significant epoch all exhibit significant ($p \leq 0.2$) deviations from zero (figure \ref{fig: Var}b), up to the point where only two earthquakes are left within the sample, above magnitude 8.25. Examining the time of greatest significance (Figure \ref{fig: Var}c), we find that this is mostly around 7 hours prior to the earthquakes. Small changes across Figure \ref{fig: Var}c likely indicate that the reduction in observed TEC is persistent for several hours at similar levels of significance as seen in Figure \ref{fig: RegVar}c.

\begin{figure}
\includegraphics[width=\textwidth]{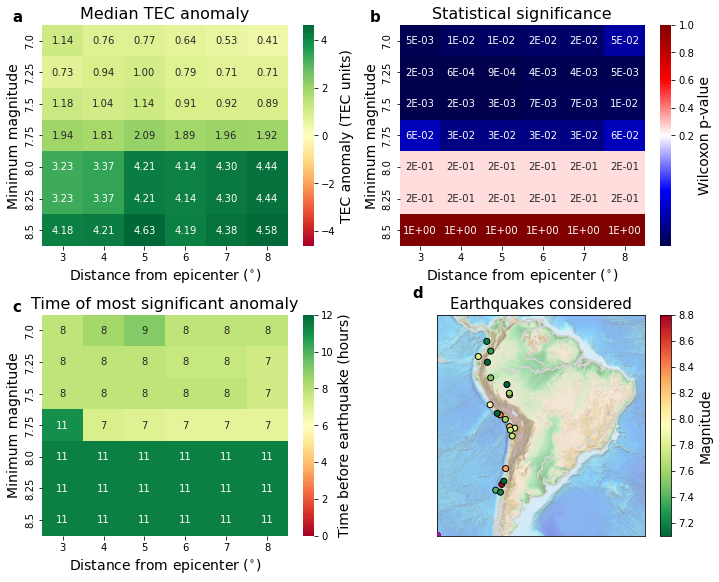}
\caption{Anomaly dependence on earthquake magnitude and distance from epicenter for South American earthquakes. Minimum magnitude implies that only events over the stated magnitude are considered. Support for magnitudes is as follows: M$>$7: 20,  M$>$7.25: 14,  M$>$7.5: 11, M$>$7.75: 6, M$>$8: 3 M$>$8.25: 3, M$>$8.5: 1. Format is similar to that in Figure \ref{fig: Var}.}
\label{fig: Var2}
\end{figure}

Figure \ref{fig: Var2} conducts the same analysis as figure \ref{fig: Var} but for the South American region which, unlike Japan, has a trend of positive TEC anomalies prior to earthquakes. Figure \ref{fig: Var2} shows the median TEC anomaly is up to $\sim1\, \mathrm{TECU}$ within $3^{\circ}$ and is typically most significant $\sim8\ $ hours prior to the event. This region supports the observations seen for the Japanese region: relatively stable TEC median anomaly values for magnitudes with sufficient significance, and an increase in TEC anomalies that can be seen further from the epicenter as magnitude increases.  A consistent time of greatest significance is observed for magnitudes with sufficient support (i.e. sufficient earthquakes within the sample). Further support for these conclusions can be drawn from all other regions with sufficient concentrations of large earthquakes (American west coast, Tongan region, Indonesian region) and are shown in the SI. 

\section{Discussion}

The results of this study suggest regionally consistent anomalies in the total electron content of the ionosphere around earthquake epicenters in the hours before major events. This section will discuss the nature of the differences between regions and briefly address possible explanations for these variations. We will then compare our results against those from previous studies before commenting on the limitations and implications of our findings. 

The key pattern seen in our results is the inconsistency of anomalies between regions. This underscores the importance of considering regional, rather than global, variations when studying pre-seismic ionospheric anomalies. Our findings are statistically significant irrespective of our choice of p-value threshold as 0.2, considering the prevalence of p-values consistently below 0.05 in Figures \ref{fig: RegVar}, \ref{fig: Var} and \ref{fig: Var2}. Given that significance remains high when considering the whole set of earthquakes above M$>$7, and averaging is avoided in favor of using quantiles, this study is robust to skewed statistics from large effects observed for only a small number of events. Local time effects are avoided through statistical analysis irrespective of event time, whose distributions are relatively uniform, suggesting that underlying regional geology is the cause of differences between regions.

Inconsistencies exist in time and polarity of TEC anomalies between regions. The Japanese subduction zone shows a negative TEC anomaly between $\sim 3$-$8$ hours prior to events, whereas the South American subduction zone shows a positive TEC anomaly $\sim6$-$10$ hours prior to an event. This statistical study is agnostic of causing mechanisms, but we must address the existence of possible explanations for these phenomena. Previous studies have hypothesized stress-activated electric currents in rocks as a mechanism for building up a positive charge beneath the Earth's surface prior to earthquakes. These build-ups of charge, known as ``positive holes'', can interact with ions in the air which in turn impact TEC measurements \cite<e.g.>[]{freund2010toward, kuo2011ionosphere}. This theory is contested amongst the pre-seismic signals community \cite<e.g.>[]{sorokin2020review, conti2021critical} and a physical explanation is not the object of this study. Nonetheless, differences in electrical signals from different rock compositions and fault types encountered in different regions could be responsible for the variation in the nature of the signal in both time and signature.

Previous studies find little to no evidence of clear precursory signatures at a global scale \cite<e.g.>[]{Thomas2017, Zhu2020, Ulukavak2020}, with analyses using a resolution of 2.5$^{\circ}$ Latitude, 5$^{\circ}$ longitude and 2 hour cadence. This conclusion is upheld with our study, but the higher resolution we have extracted from the MADRIGAL database of 1$^{\circ}$x1$^{\circ}$ spatial and 5 minute cadence allows for the identification of regionally consistent anomalies. Given the regionality of seismic properties due to underlying geology, and the regional nature of TEC anomalies we have identified, this study supports a stronger emphasis on regional breakdowns to identify local pre-seismic signals, instead of focusing on consistent global anomalies.

The statistical analysis of precursory ionospheric signals is restricted by dataset size and resolution. Globally, TEC data is only consistently available since approximately 2000, and earthquakes occur at a rate of approximately only once per month for M$>$7, and approximately once per year for M$>$8 \cite{usgs2022}. If the statistical anomalies identified in this study hold, more earthquakes in the future will further increase their statistical significance. Considering resolution, TEC data is limited by satellite and ground station coverage. As shown in Figure \ref{fig: RegVar}, in some cases there is insufficient coverage within the vicinity of earthquake epicenters to meaningfully calculate TEC variation. Furthermore the temporal resolution of 5 minute cadence for measurements may obscure, or be biased by, large variations over a shorter timescale.

The statistical nature of this study implies that although TEC anomalies are likely to occur before major earthquakes, they are not certain and do not exclusively occur before events. We envisage that the identification of TEC anomalies could be used together with other data sources in a multimodal system to help with forecasting or early warning systems for earthquake and tsunami events. To improve the regionality and resolution of our results, and create a more reliable input to forecasting models, improved resolution and real-time availability of TEC measurements are required. Future studies should evaluate the effectiveness of identifying earthquake events through precursors and focus on individual regions with maximum data availability including Japan. 

\section{Conclusion}

In this work, we have presented a data-driven exploration of ionospheric anomalies in the 12 hours before major earthquakes. We have used $1^{\circ}$x$1^{\circ}$ resolution global maps of ionospheric TEC at a five minute resolution, with a data set spanning 20 years (2000 - 2020). We define an ionospheric anomaly as the difference between the regional median of TEC observed (e.g. within $\pm 3^{\circ}$), compared to the same region 24 hours previously. We investigate if and how this anomaly might vary with time relative to earthquakes, and assess the statistical significance of any difference with the Wilcoxon test.

Through this method we have demonstrated statistically significant regional ionospheric anomalies in the 12 hours preceding earthquakes. We find that TEC is suppressed in the Japanese region between approximately 8 and 3 hours before a $M>7$ earthquake, with a median difference of $\sim 0.5\,\mathrm{TECU}$ and a Wilcoxon p-value confidence level of $p < 0.01$. In the South American region, TEC is significantly enhanced by upto $\sim 2\,\mathrm{TECU}$ between 10 and 7 hours before earthquakes with a confidence level of $p < 0.01$. Globally, we do not find a consistent, or significant ionospheric anomaly in the 12 hours prior to an earthquake.

This work has highlighted the utility of simple characterization of regional statistics, enabling studies to make robust diurnal comparisons with largely incomplete data \cite{sun2022matrix}. The significant regional differences discovered may help in identifying the debated origins of pre-seismic ionospheric anomalies. In light of our regionally variable results we suggest that future earthquake forecasting should prioritize local signals over searching for globally consistent precursors. Finally, this study has definitively shown that TEC anomalies are a valuable source of information for pre-earthquake studies that should be considered as an additional data source in earthquake forecasting models.

\section{Open Research}

The ionospheric density and seismic event datasets used for this research are open-source and available from \cite{CEDAR_2023} and \cite{usgs2022} respectively, with no restrictions.






\acknowledgments
This work has been enabled by the Frontier Development Lab Program (FDL). FDL is a collaboration between SETI Institute and Trillium Technologies Inc., in partnership with Department of Energy (DOE), National Aeronautics and Space Administration (NASA), and U.S. Geological Survey (USGS), SETI Institute, and Trillium Technologies Inc., in partnership with private industry and academia. This public/private partnership ensures that the latest tools and techniques in Artificial Intelligence (AI) and Machine Learning (ML) are applied to basic research priorities in support of science and exploration of material concerns to human kind.  The material is based upon work supported by NASA under award No(s) NNX14AT27A.

This work was also supported by UK Research and Innovation through the UKRI Centre for Doctoral Training in Application of Artificial Intelligence to the study of Environmental Risks (AI4ER).


%
\bibliography{references.bib}
%




%
%
%
%
%

\end{document}